\documentclass[11pt,twoside]{article}

\usepackage{asp2004}
\usepackage{epsf}
\usepackage{amsmath}
\usepackage{natbib}

\begin{document}
\title{Thermal Evolution of AM CVn Binary Donors}   
\author{Christopher J. Deloye and Ronald E. Taam}
\affil{Northwestern University, Dearborn Observatory, 2131 Tech Drive, Evanston, IL 60208 USA}
\author{Christophe Winisdoerffer and Gilles Chabrier}   
\affil{Ecole normale sup\'{e}rieure de Lyon, CRAL (UMR CNRS No. 5574), 69364 Lyon Cedex 07, France}    

\begin{abstract} 
We discuss results of our study on AM CVn binaries formed with donors that never ignited He before contact. For the first time, we treat the donor's in these systems in the context of a full stellar structure evolution theory and find that the binary's evolution can described in terms of 3 phases: contact, adiabatic donor expansion, and late-time donor cooling.  Details of the first and third phase are new results from this study and we focus on generally characterizing these two phases.  Finally, we present our predictions for the donor's light in these systems. 
\end{abstract}


\section*{Introduction}
The AM CVn stars are a class of interacting binaries with ultracompact orbital periods ($P_{\mathrm{orb}}$, that is $P_{\mathrm{orb}} \la 70$ min), a white dwarf (WD) accretor, and a compact donor with a He dominated composition. These systems are their progenitor systems stellar evolutionary end-state. Thus, they provide a forensic record of the binary evolution processes that formed their population. They will also be  among the gravity wave (GW) sources probed by future space borne interferometer missions such as \textit{LISA}. Determining how observations can constrain AM CVn formation processes as well as developing expectations for this population's distribution of GW signal frequencies and amplitudes depends on the details and accuracy of both our population synthesis and stellar evolution theory for these systems.

The population synthesis calculations of \citet{nelemans01a} illustrates this point.  These authors considered two formations channels for AM CVn binaries, both involving a sequence of two common envelope events that place the remnant stellar cores close enough that GW emission can drive them into contact.  The channels differ in the nature of the proto-donor prior to contact: either a non-degenerate He-burning star or a fully degenerate He WD. In their model, the formation channel determined which of two predetermined mass-radius ($M$-$R$) relations the donor is constrained to evolve along under mass loss.  Donors formed from He-stars are hotter and have larger radii, $R_2$, at fixed mass, $M_2$.  Consequently, as compared to systems with a WD donor, such systems lead to larger mass transfer rates, $\dot{M}$, at a given $P_{\mathrm{orb}}$, evolve out to longer $P_{\mathrm{orb}}$, and lead to a different distribution of GW signals.

While the \citet{nelemans01a} study utilized a \emph{single} donor $M$-$R$ relation for each formation channel, \citet{deloye05} argued that even within each channel, there should be variations in donor properties. In particular, they examined the impact of variations in the donor's entropy (or, equivalently, degeneracy) at contact for WD channel AM CVn systems.  To calculate the donor's evolution during the AM CVn phase, they utilized  set of He-object $M$-$R$ relations parameterized additionally by the donor's specific entropy, $s$, developed by \citet{deloye03} and assumed the donor's evolve adiabatically. The results were to shift the distribution of WD channel AM CVn systems off the fully degenerate track to higher $\dot{M}$ at fixed  $P_{\mathrm{orb}}$ and to shift the maximum $P_{\mathrm{orb}}$ in the population to longer periods.  In particular, some of the WD channel systems overlapped with the expected locations of He-star channel systems.

The \citet{deloye05} calculations have several shortcomings: the donor's are still constrained to evolve along predefined $M$-$R$ tracks completely defined by the donor's $s$ at contact.  These $M$-$R$ relations used were calculated assuming the donor's are fully convective and no treatment of the donor's thermal evolution was included.  It is unclear whether the assumptions of adiabatic donor structure and evolution are valid during all portions of the AM CVn phase.  These models also do not allow predictions of the donor's light.  And, finally, without full stellar models, accurate treatment of the important contact phase \citep{nelemans01a,marsh04}, during which time the $\dot{M}$ rate grows from 0 to its fully developed rate,  is not possible.  

To rectify this situation, we have calculated the evolution of WD channel AM CVn systems from pre-contact to late times treating, for the first time, the donors within a full stellar structure theory.  To do so, we developed a new Henyey-style stellar evolution code base designed to be maximally flexible in terms of input physics used and in definition of the system of ODEs to be solved.  We used the pure He equation of state developed by \citet{winisdoerffer05} that treats the difficult region of partial pressure ionization and covers the wide range of parameter space needed to follow the donor's complete evolution. We calculated $\dot{M}$ based on the prescriptions of \citet{ritter88} and \citet{kolb90}, allowing us to follow the binary's evolution during the phase of $\dot{M}$ turn-on.  Below we discuss the new insights into AM CVn evolution this improved donor treatment provides, both in terms of the binary's evolution and the donor's contribution to the binary's light. 

\section*{Phases of AM CVn Evolution}
\begin{figure}[t]
\plotfiddle{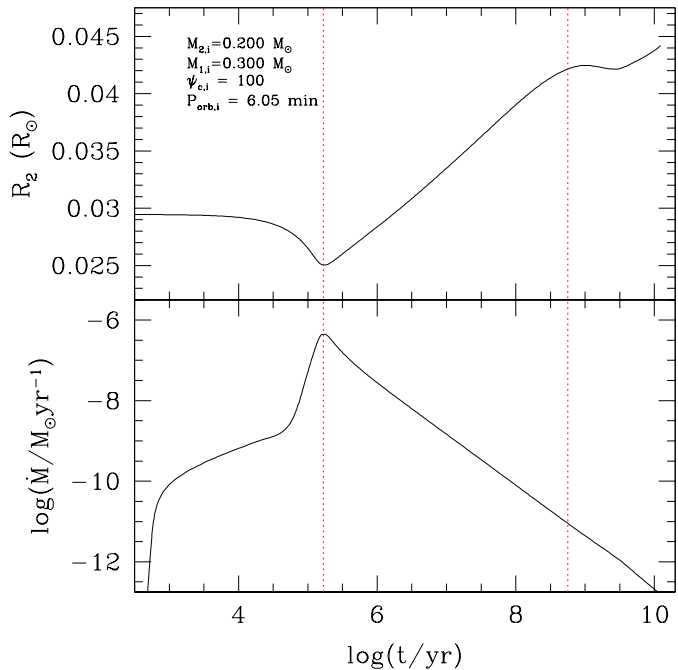}{3.15in}{0}{47.5}{47.5}{-164}{-86}
\caption{A representative result of our AM CVn binary evolution calculations showing the evolution of $R_2$ and $\dot{M}$ as a function of time. The binary's initial $M_2$, accretor mass ($M_1$), central degeneracy parameter ( $\psi_c = \varepsilon_{F,c}/k T_c$, where $\varepsilon_{F,c}$ is the electron Fermi energy at the donor's center), and $P_{\mathrm{orb}}$ at contact are indicated on the plot.  The system's evolution can be divided into three phases as marked by the dotted lines: the $\dot{M}$ turn-on phase before $\dot{M}$ maximum, a phase of adiabatic donor evolution, and a final phase of donor cooling at late times. \label{fig:deloye_f1}}
\end{figure}

We find that AM CVn evolution can be divided into 3 phases, illustrated  in Fig. \ref{fig:deloye_f1}.  The AM CVn binary's evolution initiates when the donor is brought into contact and $\dot{M}$ begins increasing from zero.  The first of our 3 phases, the $\dot{M}$ turn-on phases, consists of evolution from the start of mass transfer up to $\dot{M}$ maximum. During this phase, both $R_2$ and $P_{\mathrm{orb}}$ decrease.  By the time $\dot{M}$ reaches its maximum, $R_2$ has begun increasing and the long-lived second phase where the donor evolves nearly adiabatically in response to mass loss is underway. Finally at late times, the donor begins cooling and contracting once again, initiating the third phase of the system's evolution. The details of the first and third phases are new results, so we will focus here on providing a description of them.

The salient features of the $\dot{M}$ turn-on phase are the $R_2$ evolution and the growth rate of $\dot{M}$.  The transition from donor contraction to expansion that occurs in this phase can be explained in terms of the donor's initial $s$ profile.  As $M_2$ decreases, the pressure, $P$, at fixed mass coordinate, $m$, in the donor decrease, as does density, $\rho$, there. This results in fixed mass points moving outward in radius under mass loss.  This general trend towards expansion is counteracted by the fact that the mass lost also removes its own $R_2$ contribution.  Whether the remaining regions of the star expand enough to make up for this lost $R_2$ contribution determines whether the donor expands or contracts overall.

At fixed $m$, the contribution to $R_2$ evolution depends on $(d \ln P/dM_2)_m \sim P_c/(P M_2)$, where $P_c$ is the donor's central pressure.  The $R_2$ changes are therefore dominated by the response of the outer layers.  To determine if the donor should tend towards expansion or contraction, it is sufficient to determine $(d \ln \rho/d M_2)$ at fixed $P$ in the outer layers: if mass elements advected to lower $P$ arrive with higher $\rho$ than the element it replaces, $R_2$ will contract overall.  One can show 
\begin{equation}
 \left( \frac{d \ln \rho}{d M}\right)_P=  \frac{\chi_T}{\chi_{\rho}} \left( \nabla - \nabla' \right) \left( \frac{d \ln P}{d M_2}\right)_m\,, \label{eq:drdm_Pint}
\end{equation}
where $T$ is the local temperature,  $\nabla = (d \ln T/d \ln P)$ describes the background $T$ profile,   $\nabla'$ describes the actual $T$ evolution of the advected element, and $\chi_T/\chi_\rho = (d \ln \rho/d \ln T)_P$.

In general, $\chi_T \geq 0$, so that for $R_2$ to decrease with $M_2$, two requirements need to be met.  The first is that $\chi_T \not\approx 0$; i.e., the outer layers can not be strongly degenerate. The second is for $\nabla - \nabla' < 0$.  The latter condition occurs in radiative layers under rapid mass loss (i.e. when $\nabla' \approx \nabla_{\mathrm{ad}}$). The second condition dominates the donor's initial response; that is, initially near their surface the donors have a very steep radiative $s$ profile. Once $\dot{M}$ has increased sufficiently so that advection is nearly adiabatically, the steep $s$ gradient leads to increasing $\rho$ at fixed $P$ as much lower entropy material replaces the removed layers. This produces the overall $R_2$ decrease. The contraction is eventually reversed one the layers with a steep $s$ gradient are removed, exposing the transition to the much more shallow $s$ profile that occurs deeper in the donor's outer layers.  
\begin{figure}[t]
\plotfiddle{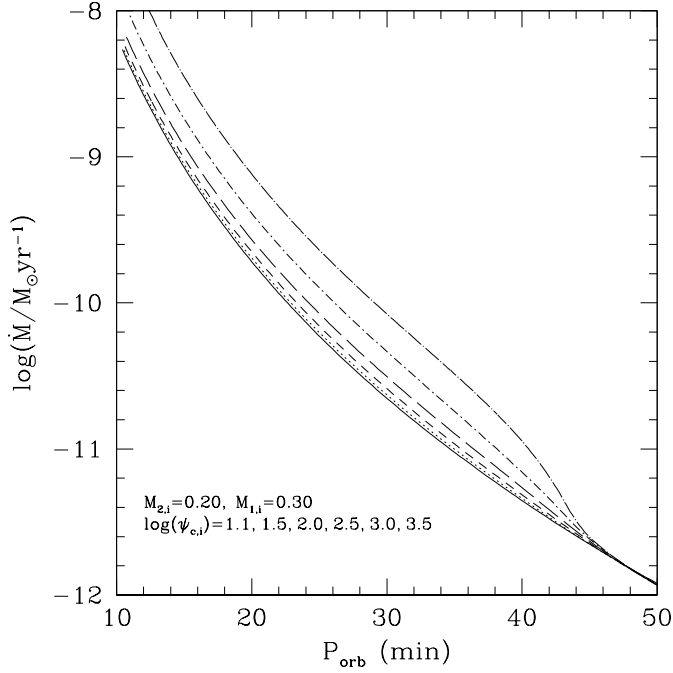}{3.15in}{0}{47.5}{47.5}{-164}{-86}
\caption{The evolution of the binary's $\dot{M}$ as a function of $P_{\mathrm{orb}}$ for systems where the initial donor and accretor masses were $0.2 M_\odot$ and $0.3 M_\odot$, respectively.  The different curves correspond to different initial donor entropies, parameterized here in terms of $\psi_{c,i}$.  The solid line shows an essentially fully degenerate object, lines above the solid line show the evolution for decreasingly degenerate donors.\label{fig:deloye_f3}}
\end{figure}

\begin{figure}[t]
\plotfiddle{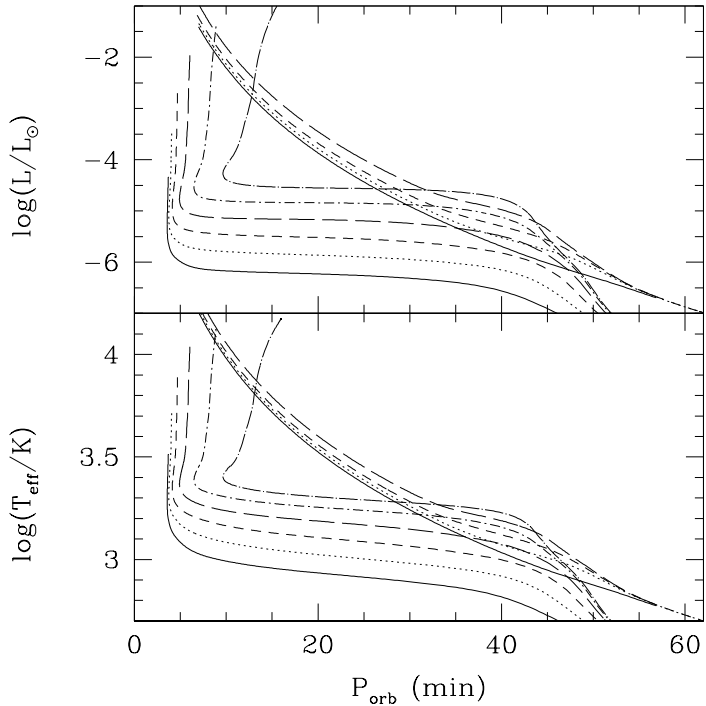}{3.4in}{0}{50}{50}{-164}{-86}
\caption{The evolution of the donor's $L$, $T_{\mathrm{eff}}$ (lower set of curves) without irradiation and of $L$, $T_{\mathrm{phot}}$ (upper set of curves) for an $\eta=0.1$. In all cases, the initial donor and accretor masses were $0.2 M_\odot$ and $0.3 M_\odot$, respectively.  The different curves correspond to the same initial donor entropies as in Fig \ref{fig:deloye_f3} with higher curves having higher entropy. \label{fig:deloye_f2}}
\end{figure}

During the turn-on phase, the $\dot{M}$ growth rate depends on the evolution of $\Delta R = R_2-R_L$. Initially, at low $\dot{M}$, $R_L > R_2$ and $\dot{M}$ grows exponentially with $\Delta R$ \citep{ritter88}. The $\dot{M}$ maximum is set by the $\dot{M}$ needed to reverse the inward evolution of orbital separation driven by GW emission.  In all cases considered, the maximum $\dot{M}$ value is greater than the $\dot{M}$ able to be supported if $R_L \geq R_2$.  Therefore $R_L$ eventually moves below the photosphere and the $\dot{M}$ growth rate slows since $R_L$ is now probing the donor's interior $\rho$ profile, not the profile of an isothermal atmosphere. The first change in the slope of $\dot{M}(t)$ seen in Fig \ref{fig:deloye_f1} is due to this effect.  The later, rapid upturn in the $\dot{M}$ growth rate is related to the evolution towards $R_2$ minimum.  The increasing $\rho$ of the donor's surface layer caused by the increasingly adiabatic advection of the underlying lower-$s$ layers produces this rapid upturn in $\dot{M}$. Less degenerate donors make contact at wider $a$, and such systems have lower $\dot{M}$ maxima since a lower $\dot{M}$ is needed to compensate for the weaker GW emission of these systems.

By the time the system has passed its $\dot{M}$ maximum, the donor is expanding adiabatically and the binary is now in what is typically thought of as the AM CVn phase.  In prior work on AM CVn evolution, the donor's adiabatic evolution has been assumed to continue indefinitely.  Our more detailed calculations show that this is not the case and, generically, the donors eventually cool and contract.  The reason for this is most easily described in terms of the evolution of two time scales: the donor's mass loss time, $\tau_M = M_2/\dot{M}$, and the donor's thermal time, $\tau_{\mathrm{th}} = c_P' T_c M_2/L$, where $c_P' = \int c_P T dm/(T_C M_2) \approx$ const. and $T_c$ is the donor's central temperature.  Once $M_2$ is much less than the accretor mass, $d \ln \tau_M/d \ln M_2 \approx 4(n-1/3)-1$. During adiabatic expansion, $d \ln \tau_{\mathrm{th}}/d \ln M_2 \approx 1 + \nabla_{\mathrm{ad},c} (2 -4 n)$ where $n = (d \ln R_2/dn \ln M_2)$ is calculated from the donor's evolution.  

For typical values of $n$ and $\nabla_{\mathrm{ad},c}$, $\tau_{\mathrm{th}}/\tau_M \propto M_2^\alpha$, where $\alpha \approx -5.5$. During the early portion of the adiabatic evolution phase, $\tau_M \ll \tau_{\mathrm{th}} $, but this inequality is eventually reversed due to this large negative $\alpha$ value.  This occurs for $P_{\mathrm{orb}} \approx 40-50$ minutes.  At this point, the donor begins shedding entropy and contracting. At late times, all evolution tracks collapse towards the fully degenerate track. This effect is illustrated in Fig. \ref{fig:deloye_f3}.

\section*{The Donor's Light}
The donor's luminosity, $L$, and effective temperature, $T_{\mathrm{eff}}$, evolution also corresponds with the donor's evolutionary phases.  As seen in Fig. \ref{fig:deloye_f2}, initially both $L$ and $T_{\mathrm{eff}}$  decrease rapidly as the system evolves toward $\dot{M}$ maximum.  Once in the adiabatic phase, both quantities plateau at a level set by the donor's initial $s$ profile.  Finally, once the donor's adiabatic evolution ends at long $P_{\mathrm{orb}}$, both $L$ and $T_{\mathrm{eff}}$ begin decreasing more rapidly again.   

We can compare the amount of accretion light the donor sees to its own luminous output.  The luminosity generated by the accretion flow is $L_{\mathrm{acc}} = \dot{M} (\phi_{L_1} - \phi_{R_1}) $, where $\phi_{L_1}$, $\phi{R_1}$ are the value of the binary's potential at the inner Lagrange point and accretors surface, respectively. At almost all times, the accretion light irradiating the donor dominates the donor's own luminous output.  To treat this effect, we performed calculations where we set the donor's outer boundary temperature to $T^4_{\mathrm{phot}}=T^4_{\mathrm{eff}} + \eta L_{\mathrm{acc}}/(16 \pi \sigma a^2)$ where $\eta$ is the overall efficiency with which $L_{\mathrm{acc}}$ is converted to photons that end up being reprocessed in the donor's atmosphere, and $a$ is the binary orbital separation. Here we define $T_{\mathrm{eff}}$ in terms of the donor's intrinsic $L$, $L = 4 \pi \sigma R_2^2 T^4_{\mathrm{eff}}$. The luminosity seen from the donor's surface, $L_s$ is then defined by $L_s = 4 \pi \sigma R_2^2 T^4_{\mathrm{phot}}$.  The evolution of $L_s$, $T^4_{\mathrm{phot}}$ for $\eta=0.1$ is shown by the upper set of lines in Fig. \ref{fig:deloye_f2}.  As compared to the non-irradiated case, $L_s$ can be increased significantly over $L$ for a large portion of the AM CVn phase.  Overall, with $\eta=0.1$, we expect donors to have $L_s$ in the range of $10^{-6}$-$10^{-3} L_\odot$ and $T_{\mathrm{phot}}$ between 1000-5000 K for $P_{\mathrm{orb}} < 50$ minutes. Both quantities will be higher if $\eta$ is larger.


\acknowledgements 
We would like to thank Dean Townsley for helpful discussions related to the preparation of this contribution.  This work was supported in part by the NSF through grant AST 02-00876. 



\end{document}